\documentstyle[12pt]{article}
\begin{document}
\title{Galaxies and Superclusters}
\author{B.G. Sidharth$^*$\\
Centre for Applicable Mathematics \& Computer Sciences\\
B.M. Birla Science Centre, Adarsh Nagar, Hyderabad - 500 063 (India)}
\date{}
\maketitle
\footnotetext{$^*$Email:birlasc@hd1.vsnl.net.in; birlard@ap.nic.in}
\begin{abstract}
In this brief note, we would like to point out that large-scale structures
like galaxies and superclusters would arise quite naturally in the universe.
\end{abstract}
There being about $N = 10^{11}$ galaxies in the universe, each about
$l = 10^{23}cms$ across, we can easily verify that
\begin{equation}
R \sim \sqrt{N}l\label{e1}
\end{equation}
where $R$ is the radius of the universe.\\
As is well known, (\ref{e1}) arises in the theory of Brownian motion on the
one hand\cite{r1} and on the other, it is also true if $N$ represents the
number of elementary particles $\sim 10^{80}$, in the universe and $l$, their
size or spread, that is their Compton wavelength\cite{r2}.\\
We can now interpret (\ref{e1}) as follows:\\
From the large-scale perspective, the galaxies are approximately in Brownian
motion and their size is given correctly by (\ref{e1}). Interestingly, as we
can easily verify, an identical relation holds for superclusters also\cite{r3},
with a similar interpretation.\\
Moreover (\ref{e1}) implies a two dimensional structure - this is indeed true;
not only do galaxies have large flat disks, but also superclusters have a flat
cellular character\cite{r4}.\\
However, (\ref{e1}) is not true for stars. In this case gravitation is strong
and the Brownian approximation is no longer valid.\\
Thus galaxies and superclusters would naturally arise in the Universe.

\end{document}